\documentclass[a4paper]{VisionStyle}
\usepackage{epsfig}

\def\ecs{ergs cm$^{-2}$ s$^{-1} \ $}

\begin{document}

\title{The XMM-Newton Bright Serendipitous Source Sample (XMM-Newton BSS)}

%\author{G.L.\,Pilbratt\inst{1} \and F.\,Favata\inst{1} \and 
%  M.A.N.Y.\,Unknowns\inst{2} } 

\author{
R.\,Della Ceca\inst{1} \and 
T.\,Maccacaro\inst{1} \and 
A.\,Caccianiga\inst{1} \and 
P.\,Severgnini\inst{1} \and
X.\,Barcons\inst{2} \and
D.\,Barret \inst{3} \and
F.\,Carrera\inst{2}  \and 
G.\,Hasinger \inst{4} \and
R.G.\,McMahon  \inst{5} \and
C.\,Motch  \inst{6} \and
W.\,Pietsch \inst{4} \and
M.J.\,Page \inst{7} \and
S.\,Rosen \inst{7} \and
A.\,Schwope \inst{8} \and
M.G.\,Watson \inst{9} \and
N.A.\,Webb \inst{3} \and
D.M.\,Worrall  \inst{10} \and
H.\,Ziaeepour  \inst{7}
}
\institute{
  Osservatorio Astronomico di Brera, via Brera 28, 20121, Milano, Italy 
\and 
Instituto de F\'\i sica de Cantabria (CSIC-UC), 39005 Santander, Spain
\and
Centre d'Etude Spatiale des Rayonnements, 9 Avenue du Colonel Roche, 31028 Toulouse Cedex 04, France
\and
Max-Planck-Institut f\"ur Extraterrestrische Physik, Postfach 1312, 85741 Garching, Germany 
\and
Institute of Astronomy, Madingley Road, Cambridge CB3 0HA, UK 
\and
Observatoire Astronomique de Strasbourg, 11 rue del'Universit\'e, 67000 Strasbourg, France 
\and
 Mullard Space Science Laboratory, UCL, Holmbury St Mary, Dorking, Surrey RH5 6NT, UK 
\and
Astrophysikalishes Institut Potsdam, An der Sternwarte 16, 14482 Potsdam, Germany 
\and  
Department of Physics and Astronomy, University of Leicester, LE1 7RH, UK
\and
Department of Physics, University of Bristol, Royal Fort, Tyndall Avenue, Bristol, BS8 1TL, UK
}

\maketitle 

\begin{abstract}

The XMM-Newton Survey Science Center is constructing  a large 
($\sim$ 1000 sources) and complete sample
of bright serendipitous XMM-Newton sources at high galactic latitude, so as 
to allow both the discovery of sources of high individual interest as well 
as statistical population studies.
The sample, known as ``The XMM-Newton Bright Serendipitous Source Sample" 
(XMM-Newton BSS), has a flux limit of $\sim 10^{-13}$ erg cm$^{-2}$ s$^{-1}$ 
in the 0.5--4.5 keV energy band and will be fundamental 
in complementing other medium and deep survey programs 
(having fluxes 10 to 100 times fainter) in order to:
a) characterize the  X-ray sky and its constituents,  
b) understand the evolution of the population selected and 
c) define statistical identification procedures to select 
rare and interesting populations of X-ray sources (e.g. 
high z clusters of galaxies, BL Lacs, type 2 QSOs, etc..).
We discuss here the scientific rationale and the 
details of the project and we present some preliminary 
results.
\end{abstract}

\section{Introduction and Scientific Rationale}
  
X-ray selection provides a well-proven and efficient mean of  finding some of
the astrophysically most interesting and energetic objects in the 
Universe, like Active Galactic Nuclei (AGN) and 
clusters of galaxies. 

The XMM-Newton Survey Science Center (SSC)
\footnote {
The XMM-Newton SSC is an international collaboration involving a consortium of 
several institutions. The XMM-Newton SSC was appointed by  
ESA to exploit the XMM-Newton serendipitous 
detections for the benefit of the international scientific community
(\cite{rdellaceca-WB1:Watson2001}). 
See http://xmmssc-www.star.le.ac.uk for a full description of the program
and the Barcons et~al., contribution at this meeting for an updated status 
of the XMM-Newton XID program.}
is assembling a large (about 1000 sources) sample
of bright serendipitous XMM-Newton sources at high galactic latitude ($|b| > 20^o$), 
following 
well defined criteria (completeness, representativeness, etc..) so as 
to allow both a detailed study of sources of high individual interest as well 
as statistical population studies.
\par\noindent
The sample, known as ``The XMM-Newton Bright 
Serendipitous Source Sample" 
(XMM-Newton BSS), 
is designed to have a flux limit of $\sim 10^{-13}$ erg cm$^{-2}$ s$^{-1}$ 
in the 0.5--4.5 keV energy band.
The major scientific goals of this project are to study 
the mix of different kinds of AGN, and 
the cosmological evolution of clusters of galaxies.

AGN are expected to be the most numerous class of X-ray sources.
The majority of the XMM-Newton BSS sources will 
have enough statistics (hundreds of counts) to allow X-ray spectral 
investigations (see e.g. \cite{rdellaceca-WB1:Dellaceca2000} for the discovery of a Type 2 
AGN from X-ray data only). 
Such X-ray spectral information is fundamental to determine, source-by-source, 
the amount of metal absorption, the spectral shape, and the presence of reprocessed 
features (e.g. FeK line). 
Moreover, at the X-ray limiting flux of the XMM-Newton BSS sample the optical counterparts
of the X-ray sources have magnitudes bright enough to allow relatively 
high-quality optical and near-infrared spectroscopy even with a 4 meter 
class telescope.
X-ray data, combined with optical and near-infrared spectroscopic information, 
will allow us to investigate the poorly understood relationships
(e.g., \cite{rdellaceca-WB1:Maiolino2001}) between narrow- and broad-line AGN in
the optical regime, and absorbed and unabsorbed AGN in the X-ray regime.
The completeness of our sample should place our investigations on firm
and solid statistical ground.  The large area covered by the XMM BSS
will allow us also to investigate in detail rare 
source population (e.g. the long sought type 2 QSOs).

Another crucial issue for cosmology is the evolution of the XLF 
of clusters of galaxies. 
Early claims of a negative evolution with redshift are still being 
tested and refined.
For example, the ROSAT results from the RDCS survey 
(e.g. \cite{rdellaceca-WB1:Rosati1998};
\cite{rdellaceca-WB1:Borgani2001})
show that the negative 
evolution, if any, is mostly concentrated on the high luminosity 
($L_x > 5\times 10^{44}$ erg s$^{-1}$) part of the XLF. 
About 120 (in case of negative evolution)  
and 160 (in case of no evolution) clusters are expected in the 
XMM-Newton BSS. The largest discrepancy is on the predicted number of high-luminosity
($L_x > 5\times 10^{44}$ erg s$^{-1}$) systems: 
$\sim$ 45 are expected in the no evolution case while $\sim$ 15 
are expected in the negative evolution case.
These numbers, although indicative, illustrate the power of the 
XMM-Newton BSS as a tool to solve a long-standing problem of the 
modern physical cosmology.
Medium and deep surveys, covering a smaller area of the sky, 
will sample more efficiently the low luminosity part of the 
XLF which is already well sampled by the RDCS and for which 
no strong evidence of cosmological evolution is found 
(e.g. \cite{rdellaceca-WB1:Rosati1998}).
We stress here that 
the search for rare high-luminosity clusters is best done  
in shallow surveys covering a large area, such as the XMM-Newton BSS.

\smallskip

When combined with
the results from other XMM-Newton medium and deep survey programs
(in the same energy band,  e.g. AXIS, 
\cite{rdellaceca-WB1:Barcons2002}; 
or in different energy selection bands, e.g.  
HELLAS2XMM - \cite{rdellaceca-WB1:Baldi2002}; 
\cite{rdellaceca-WB1:Hasinger2001})
the XMM-Newton BSS will extend the baseline for all evolutionary studies 
(as e.g. the AGN and cluster of galaxies luminosity function) as well as 
it will allow us to investigate selection effects in the source population due 
to the used energy selection band.
The importance of a large baseline in luminosity and  
redshift has been already shown by 
\cite{rdellaceca-WB1:Miyaji2000} who have 
combined several surveys of different depth and solid angle carried out with 
ROSAT to investigate in deeper detail than previously possible the cosmological 
properties of the AGN population.

An important part of our work with the XMM-Newton BSS will be to define 
statistical identification procedures to select
rare and interesting  populations of X-ray sources,  
enabling the application of these procedures to the vast amount of
XMM-Newton serendipitous data that will be accumulated during the lifetime of
the mission.

%Finally, through the XMM-Newton BSS it will be possible
%to define statistical identification procedures to select
%rare and interesting  populations of X-ray sources  
%applying these procedures to the ``huge" 
%amount of XMM-Newton data that will be accumulate in the next few years.

This paper is organized as follows.
In section 2 we describe the strategy used to define the source sample.
Preliminary results from this project are presented in Section 3, while the summary and 
conclusions are reported in Section 4.

\section{Survey Strategy}
%\label{fauthor-E1_sec:tit}

The bright catalogue is designed to contain about 1000 sources with an X-ray
flux limit in the 0.5--4.5 keV energy band of $\sim$10$^{-13}$  erg s$^{-1}$
cm$^{-2}$. The size of the sample is dictated by the need 
to reveal and to characterize statistical properties  also 
for minority populations (e.g. high-z clusters
of galaxies, BL Lacs objects etc..).

\subsection{Why the 0.5-4.5 keV energy band ?}
The choice of the 0.5--4.5 keV energy band is motivated 
by the desire to avoid the very soft photons (minimizing 
non-uniformities introduced by the different values of 
Galactic absorbing column densities along the line of 
sight) and by the need of compromising between a 
broad passband (to favour throughput) and a narrow passband 
(to minimize non-uniformities in the selection function due to 
different source spectra).
Furthermore in the 0.5--4.5 keV band XMM-Newton has the highest throughput.

\subsection{Which primary selection camera?}

We have decided to use  only data from the EPIC MOS2 detector 
to define the source sample.
The main reasons for this choice are:
\begin{enumerate}

\item unlike the EPIC pn, the EPIC MOS detectors have a circular symmetry  that
simplifies the analysis of the field. 
For example, in the case of the EPIC MOS detectors the
source target is, in the large majority of the  observations, in the 
central chip;

\item the PSF in the 2 EPIC MOSs is narrower than the PSF in the EPIC pn.
In particular, the EPIC MOS2 has the narrowest PSF 
(FWHM$\sim$4.4$^{\prime\prime}$ and
HEW$\sim$13.0$^{\prime\prime}$  at 1.5~keV, see \cite{rdellaceca-WB1:Ehle2001}).  
As a comparison, the EPIC pn PSF has FWHM$\sim$6.6$^{\prime\prime}$  and
HEW$\sim$15.2$^{\prime\prime}$;

\item the dead spaces between the EPIC MOS chips are narrower than
the gaps in the EPIC pn detector, simplifying source detection and 
analysis and maximizing the search area;

\item unlike with the EPIC pn camera, we can use the EPIC MOS2 observations in large- 
and small-window mode excluding the area occupied by the central chip. 
Since $\sim$ 30\% of the observations 
have been performed in window mode, retaining these 
observations  will maximize the searched area,  
speeding up the creation and definition of the source sample.

\end{enumerate}

The major disadvantage of the EPIC MOS2 camera when compared to
the EPIC pn camera is the reduced sensitivity.
The effective area of EPIC MOS is smaller than that of EPIC pn, and 
this is particularly true
at high  (E$>$5 keV) and  low (E$<$1.5 keV) energies 
(note that for the selection band adopted, 0.5--4.5 keV, only
the reduced effective area at  E$<$1.5 keV is important). 
However, since the sample contains relatively bright sources,  
and considering a minimum exposure time of 5000 s (see below) 
this lower efficiency {\bf does not affect the source selection}.
Obviously, once the source is detected and included in the sample, 
additional information using data from the EPIC MOS1 and pn 
detectors can be collected in order to increase the statistics
for the X-ray spectra and  morphology analysis.

\subsection{Criteria for field selection}

Not all the available EPIC MOS2 pointings are adequate for producing the
bright source catalogue. We have defined a set of selection criteria in order 
to avoid problematic regions of the sky,  
to maximize the availability of ancillary information at 
other frequencies (i.e. optical and radio) and to  
speed up the optical identification process.  
The fields will be thus selected on the basis of the following
criteria:

\begin{enumerate}

\item availability to SSC (XMM-Newton fields with PI granted permission
\footnote{This restriction of course only applies during the 
proprietary period of the XMM-Newton observations.}
plus 
XMM-Newton public fields);

\item  high Galactic 
latitude ($|b| \geq$20$^{\circ}$) to avoid crowded fields, to obtain a relatively
``clean" extragalactic sample  
and to have magnitude information for the optical counterparts 
from the Digital Sky Survey (DSS) material
(the Automated POSS Machine - APM - catalogue is almost
complete for  $|b| \geq$20$^{\circ}$);

\item galactic absorbing column density along the line of 
sight less than 10$^{21}$ cm$^{-2}$, to minimize
non-uniformities introduced by large values of the 
Galactic $N_H$;

%\item {\bf $\delta\geq$--40$^{\circ}$}: to match the NRAO VLA Sky Survey 
%(NVSS) sky-coverage. In this way all the sources in the  
%sample will have a radio flux density and a good VLA position if they
%are brighter than $\sim$2.5 mJy;

\item exclusion of fields centered on bright and/or extended X-ray targets 
and/or  containing very bright stars. 
In the first case the effective area of sky covered and 
the actual flux limit are difficult to estimate correctly,  making
the derivation of the sky-coverage more uncertain; in the latter case 
the search for the optical counterpart of the X-ray sources 
could be very difficult or even impossible due to the
presence of the bright star;

\item good time interval (GTI) exposure $>$ 5ks. With this constraint all the 
sources with a count rate (0.5 - 4.5) keV $\geq$ 0.01 cts/s will be 
detectable across the whole field of view considered 
ensuring a flat sensitivity;

\item finally, we have also excluded EPIC MOS2 pointings 
suffering from a high background rate (i.e. 
accumulated during particle background flares). 

\end{enumerate}

Note that we have also considered the EPIC MOS2 observations in large- and small-window 
mode satisfying the criteria discussed above; in these cases
we exclude the central chip from the 
investigated area but use the area covered from the other six CCDs.

\subsection{Source detection}

The EPIC MOS2 observations appropriate for this project have been processed through 
the pipeline processing system, using tasks from the XMM-Newton Science Analysis 
Software.
Full details about the pipeline processing system, the pipeline products
as well as the source searching procedures, flux measurements, etc..
can be found in \cite{rdellaceca-WB1:Watson2001} 
and http://xmmssc-www.star.le.ac.uk.
 
\subsection{Criteria for source selection}

Having decided the field-selection criteria, 
we define here the criteria for the source selection within each EPIC MOS2 field:

\begin{enumerate}

\item 0.5--4.5 keV count-rate $\geq$0.01 cts/s. 
This count-rate, already corrected for 
vignetting and PSF corresponds, for a
source with a power-law spectrum with energy index $\alpha_E$=0.9 (see Section 3.3),
to a flux (0.5 - 4.5 keV) of 
$\sim 7\times$10$^{-14}$ erg s$^{-1}$ cm$^{-2}$; 
%We note that for a minimum exposure time of 5000 sec a source 
%at the flux limit will deposit in the detector more than 50 net 
%counts, allowing a firm detection of the source and ensuring a flat 
%sensitivity across the field (i.e. flat sky coverage).  

\item sources with a distance from the EPIC MOS2 center
between an inner radius ($R_{in}$) and an outer radius
($R_{out}$). $R_{in}$ depends  on 
a) the actual size and brightness of the target and 
b) on the window mode. 
$R_{in}$ ranges between 0 (e.g. survey fields with no ``target") and 8 arcmin
(bright/extended X-ray sources and/or large- and small-window mode).
In this way the area of the detector ``obscured'' by the presence
of the target and/or not exposed is excluded from the analysis.
$R_{out}$ is, for the large majority of the fields, equal to 
13 arcmin; 

\item we have also excluded the sources too close to the edges of the 
field of view 
or to the gaps between the CCDs. 
We note that these sources could have either the flux and/or the 
source centroid affected by the proximity to the 
edges and/or the gaps, and therefore could represent a problem 
in the subsequent analysis and/or interpretation of the data.
The excluded area has been taken into account in the computation of the sky 
coverage;

\item finally, in order to guarantee that all the sources in the catalogue 
are truly serendipitous we have excluded the target and the sources physically 
related to the target. 

\end{enumerate}

\section{Preliminary Results}
%\label{fauthor-E1_sec:cmd}

At the time of this writing (January, 2002) 104 suitable XMM-Newton fields have been already analyzed 
and a  first sample of 185 sources selected. 
Some general considerations, tested and 
verified using this initial sample, are in order:
\begin{itemize}
\item the majority of the X-ray sources have enough statistics
(hundreds of counts when data from EPIC MOS1 and pn are also 
considered) to allow X-ray studies in terms of
energy distribution, source extension and flux variability; 
\item the optical counterpart,  
for the large majority (85\% - 90\%) of the objects, has a magnitude above the 
POSS II limit (R $\sim 21^{mag}$), thus allowing spectroscopic 
identification at a 4m class telescope.
Furthermore, given the positional accuracy of XMM-Newton for the bright sources
(90\% error circle of 2-5 arsec) and the 
magnitude of the optical counterparts, only one object needs to be 
observed to secure the optical identification. 
Among the blank fields, rare and interesting 
classes of X-ray sources (e.g. high z clusters of galaxies, BL Lacs, 
highly absorbed AGN) could be present;

\item about $20\%$ of the sources are identified through a literature 
search (NED, SIMBAD);

\item for all the sources north of $-40^o$ radio information is available 
down to a few mJy level (NRAO VLA Sky Survey, NVSS; \cite{rdellaceca-WB1:Condon1998}). 
Information at other wavelength, although incomplete will be 
abundant (e.g. the 2MASS survey, 
optical colors). 
\end{itemize}

\subsection{The number-counts relationship}

\begin{figure}[ht]
  \begin{center}
    \epsfig{file=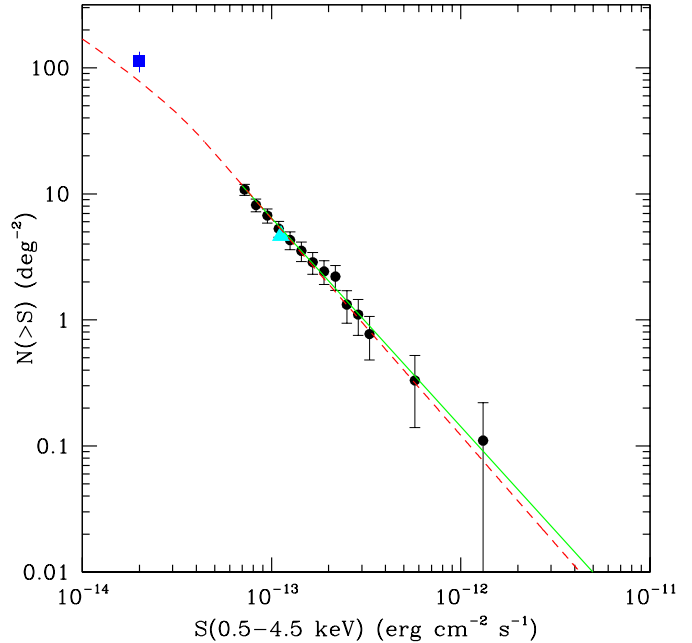, width=9cm}
  \end{center}
\caption{The extragalactic number-flux relationship obtained using 
the XMM-Newton BSS sample assembled so far (binned representation:
black filled circles)
The Log(N$>$S)-LogS is well described by a power-law model having a 
slope equal to $1.65 \pm 0.19$ (green solid line).
In the same figure we have also reported the ROSAT (0.5 -- 2.0 keV) 
Log(N$>$S)-LogS 
(dashed red line) and the EMSS (0.3--3.5 keV) extragalactic number density 
(sky-blue triangle) both converted to the 0.5--4.5 keV energy band 
using a power-law spectral model having $\alpha_E = 0.9$.
The blue filled square at $S \simeq 2\times 10^{-14}$ \ecs
represents the extragalactic surface density obtained by the 
XMM-Newton AXIS Medium Survey team (Barcons et~al., 2002).  
}  
\label{rdellaceca-WB1_fig:fig1}
\end{figure}

In Figure~\ref{rdellaceca-WB1_fig:fig1} we show (filled circles) a binned  
representation of
the number-flux relationship, obtained by folding the  sky
coverage with the flux of each source.
As already said, the sky coverage of the XMM-Newton BSS  
at the flux limit used to define the sample, is flat 
and up to now $\sim 9.1$ sq.deg.  have been covered. 
A conversion factor appropriate for a 
power-law spectral model with energy index equal to 0.9
(see Section 3.3), 
filtered by an $N_{H_{Gal}} = 3\times 10^{20}$ cm$^{-2}$,   
has been used in the conversion between the count rate and the 
flux [1 cts/s (0.5--4.5 keV) = $7.2 \times 10^{-12}$ \ecs 
(0.5--4.5 keV)].  
This count rate to flux conversion factor is accurate 
$\pm 20\%$ for all the energy spectral indices in the 
range between 0.0 and 2.0 and for all the $N_{H_{Gal}}$ 
in the range between $\sim 10^{20}$ cm$^{-2}$
and $\sim 10^{21}$ cm$^{-2}$.

Because we are primarily interested in the extragalactic number-flux 
relationship 
we have excluded from the computation the sources classified as
(or suspected to be) stars.
In the same figure we also report a parametric representation (green solid
line) of the LogN($>$S)--LogS obtained by applying the maximum
likelihood method to the unbinned data (see 
\cite{rdellaceca-WB1:Gioia1990} for details).
The fit has been performed from a flux of $7.2\times 10^{-14}$
\ecs (the faintest flux) to a flux of $\sim 10^{-12}$ \ecs.
For fluxes brighter than this limit we may not be complete since most
of the ``bright" X-ray sources were chosen as target of the
observation and then excluded, by definition, from the survey.
However we note that the space density of sources with flux greater
than $\sim 10^{-12}$ \ecs is such that fewer than 2 sources are
expected given the sky coverage area.  

The LogN($>$S)--LogS can be described by a power-law model 
N($>$S) = K $\times S^{-\alpha}$ with best-fit value
for the slope $\alpha = 1.65\pm 0.19$ ($68\%$ confidence interval). 
The normalization K is determined by rescaling the model to the actual
number of objects in the sample and, in the case of the ``best" fit
model, is equal to K = $2.21 \times 10^{-21}$ deg$^{-2}$.  

The red dashed line represents the extragalactic 
ROSAT (0.5 -- 2.0 keV) 
Log(N$>$S)-LogS adapted from \cite{rdellaceca-WB1:Hasinger1998} and 
converted to the 0.5--4.5 keV energy band 
using a power-law spectral model having $\alpha_E = 0.9$.
The sky-blue filled triangle at $S \simeq 1\times 10^{-13}$ \ecs
represents the 
extragalactic number density at the  flux limit 
of the EMSS \cite{rdellaceca-WB1:Gioia1990}
converted from the original (0.3--3.5 keV) to 
the  0.5--4.5 keV energy band using the same spectral model as before.
Finally, the blue filled square at $S \simeq 2\times 10^{-14}$ \ecs
represents the extragalactic surface density obtained from the 
XMM-Newton AXIS Medium Survey team in the same energy range of the 
XMM-Newton BSS (\cite{rdellaceca-WB1:Barcons2002}).  

As the figure clearly shows, a very good agreement with previous and 
new results is obtained assuring us about data analysis and selection.
Consistent results are also obtained from a comparison 
with the HELLAS2XMM survey (\cite{rdellaceca-WB1:Baldi2002}).
%{\bf check}. 

\subsection{Optical identification and classification}

The XMM-Newton BSS sample assembled so far is composed of 185 sources.
Up to now 59 source have been spectroscopically identified; 39 of these
identifications come from the literature (NED, SIMBAD) while 20 from the 
XMM-Newton XID AXIS project (\cite{rdellaceca-WB1:Barcons2002}).
The optical breakdown is the following: 30 broad line AGN (QSOs/Seyfert 1),
3 type 2 AGNs,  1 optically ``normal" galaxy, 5 clusters of galaxies, 
1 BL Lac object and 19 stars. 
However we stress that this sample 
of identified objects is probably not representative of the whole population.
A flavour of the dominant X-ray source population in the 
0.5--4.5 keV energy range at an X-ray flux limit of 
$S \simeq 2\times 10^{-14}$ \ecs may be obtained from the 
almost complete (93\%) spectroscopic identification of a small sample of  
29 X-ray sources selected from two XMM-Newton fields studied in the 
AXIS project (\cite{rdellaceca-WB1:Barcons2002}).

\subsection{Spectral properties}

A ``complete" spectral analysis for all the sources in the bright sample
(using data from the two EPIC MOSs and the EPIC pn) is in progress. 
In the meantime and in order to gain X-ray spectral information 
we present here a ``Hardness Ratio" analysis of the 
single sources using only EPIC MOS2 data;
this latter method is equivalent to the ``color-color" analysis 
largely used at optical wavelengths. 

We have used the hardness ratios as defined from the XMM-Newton pipeline 
processing: 
~~~~~~\\
 \[
HR2={C(2-4.5\, {\rm keV})-C(0.5-2\, {\rm keV})\over C(2-4.5\, {\rm
keV})+C(0.5-2\, {\rm keV})}
\]
and
\[
HR3={C(4.5-7.5\, {\rm keV})-C(2-4.5\, {\rm keV})\over C(4.5-7.5\, {\rm
keV})+C(2-4.5\, {\rm keV})}
\] 
~~~~~~\\
where C(0.5$-$2\, {\rm keV}), C(2$-$4.5\, {\rm keV}) and 
C(4.5$-$7.5\, {\rm keV}) are the count rates
\footnote{Note that the count rates have been computed using the 
exposure-map in each energy band, so that energy dependent 
vignetting is approximately corrected.}
in the 0.5$-$2, 2$-$4.5 and 4.5$-$7.5 keV energy bands, respectively.
We have not used here HR1 which is defined using 
C(0.15$-$0.5\, {\rm keV}) and C(0.5$-$2\, {\rm keV}) energy band since 
the count rate in the  (0.15$-$0.5\, {\rm keV}) energy band is a strong 
function of the Galactic absorbing column density along the line of sight.
Note that the effect on HR2 and HR3 due to the different $N_{H_{Gal}}$ 
for the objects in the sample (which range between $\sim 10^{20}$ to 
$10^{21}$ cm$^{-2}$) is completely negligible.

\begin{figure}[ht]
  \begin{center}
    \epsfig{file=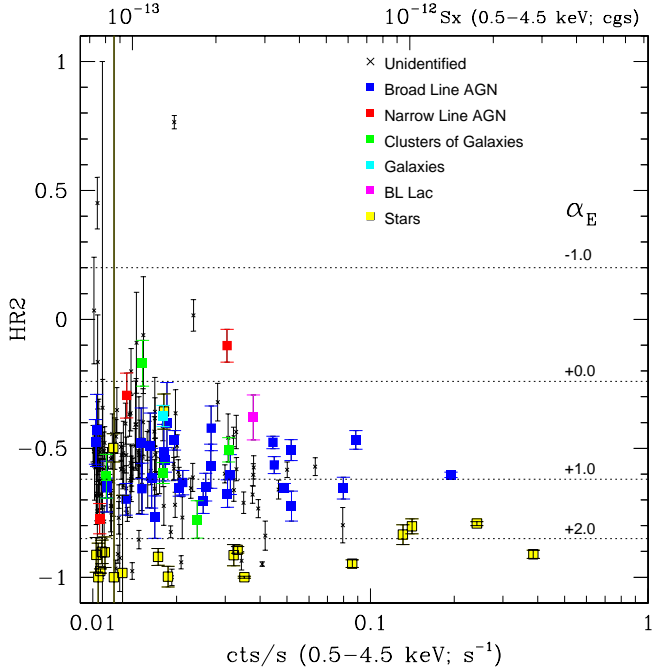, width=9cm}
  \end{center}
\caption{HR2 vs. EPIC MOS2 count rate in the (0.5--4.5) keV energy band for the
185 objects assembled so far, compared with the 
HR2 expected from a non-absorbed power-law model with $\alpha_E$ ranging 
from -1 to 2 ($S_E \propto E^{-\alpha_E}$).
For comparison the energy spectral index of the CXB ($\alpha_E = 0.4$) 
corresponds to HR2$=-0.39$.
The flux scale on the top has been computed assuming a conversion factor 
appropriate for $\alpha_E = 0.9$, the mean energy spectral index 
of the ``extragalactic" sample in the (0.5--4.5) keV energy band.
We have used different colors to mark the identified objects.} 
\label{rdellaceca-WB1_fig:fig2}
\end{figure}

In Figure~\ref{rdellaceca-WB1_fig:fig2} we show the HR2 as a function 
of the count rate in the 
(0.5 -- 4.5) keV energy band.
On the top, we have reported 
the flux scale computed assuming a conversion factor 
appropriate for $\alpha_E  = 0.9$, which is the mean energy spectral index 
of the ``extragalactic" sample in the 0.5--4.5 keV energy band (see below).
We have used different colors to mark the spectroscopically identified and 
classified objects.  
It is worth noting that 
a) all but 2 ($\sim 90\%$) of the sources classified as stars have an HR2 
less than $-0.8$ and b) none of the X-ray sources identified so far with 
extragalactic objects have HR2 below $-0.8$. 
If we assume a simple Raymond-Smith thermal model, HR2$\leq -$0.8 
corresponds to temperatures below $\sim 1.5$ keV, in very good 
agreement with the stellar identification.
This is a first nice result on the 
statistical broad band properties of the sample; this criterion, 
if confirmed from better statistics, could be used to select (or exclude 
depending on the scientific goal) the stars from the XMM-Newton source database.
We are at the moment investigating the combination of this criteria 
with the usual X-ray to optical flux ratio.

No obvious trend of the source spectrum as a function of the count rate
is clearly visible. If we exclude the objects classified as stars 
the weighted average is HR2$=-0.58\pm 0.02$ corresponding to 
$\alpha_E = 0.9\pm 0.1$. For comparison the 
weighted average of the sources classified as Broad Line AGN, 
clusters of galaxies and  unidentified objects is 
HR2$=-0.56\pm 0.02$, HR2$=-0.52\pm 0.1$ and 
HR2$=-0.58\pm 0.03$, respectively.
Two of the three sources identified as Narrow Line AGN seem to have 
an energy spectral index flatter than that of the cosmic X-ray 
background ($\alpha_E = 0.4$ corresponding to HR2$=-0.39$); 
on the contrary none of 
the X-ray sources identified as Broad Line AGN seem to have a spectra 
flatter than $\alpha_E = 0.4$.
Finally among the unidentified sources  about 13\% seem to be described 
by an energy spectral index below $\alpha_E = 0.4$.

\begin{figure}[ht]
  \begin{center}
    \epsfig{file=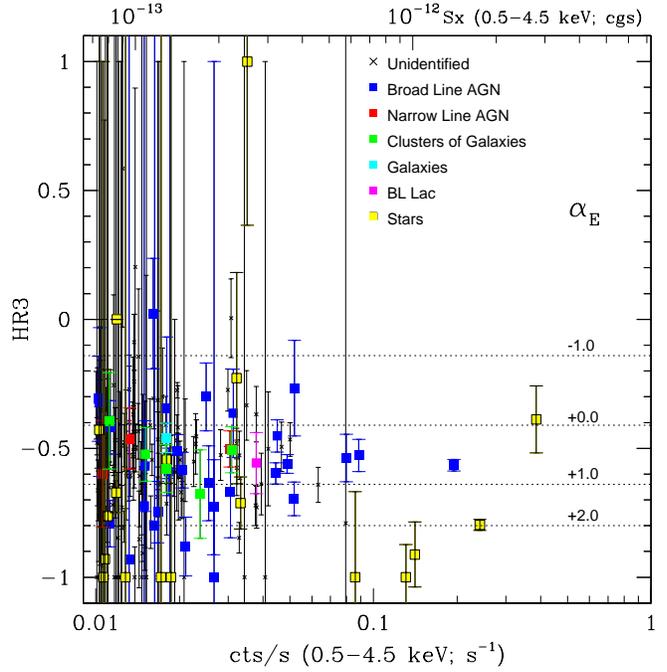, width=9cm}
  \end{center}
\caption{HR3 vs. EPIC MOS2 count rate in the (0.5--4.5) keV energy band for the
185 objects assembled so far, compared with the 
HR3 expected from a non-absorbed power-law model with $\alpha_E$ ranging 
from -1 to 2 ($S_E \propto E^{-\alpha_E}$).
For comparison the energy spectral index of the CXB ($\alpha_E = 0.4$) 
corresponds to HR3$=-0.51$.
The flux scale on the top has been computed assuming a conversion factor 
appropriate for $\alpha_E = 0.9$, the median energy spectral index 
of the sample in the (0.5--4.5) keV energy band.
We have used different colors to mark the identified objects. 
}  
\label{rdellaceca-WB1_fig:fig3}
\end{figure}

In Figure~\ref{rdellaceca-WB1_fig:fig3} we report HR3 as a function 
of the count rate in the 
(0.5 -- 4.5) keV energy band (symbols and flux scale on the 
top are as in Figure~\ref{rdellaceca-WB1_fig:fig2}).
This hardness ratio is much noisier since many of the sources are 
detected with poor statistics (or even undetected) in the 4.5--7.5 keV
energy band. No separation between the different classes of 
X-ray sources is clearly visible although the large errors could 
prevent us from finding possible differences. 
With the same caveat, no obvious trend of the source spectra as a 
function of the count rate is clearly visible; the median 
HR3 of the ``extragalactic" population is 
HR3$\sim -0.57$ corresponding to $\alpha_E \sim 0.7$.
%HR3$\sim -0.57$
%is also the median HR3 of the sources classified as Broad Line AGN.

\begin{figure}[ht]
  \begin{center}
    \epsfig{file=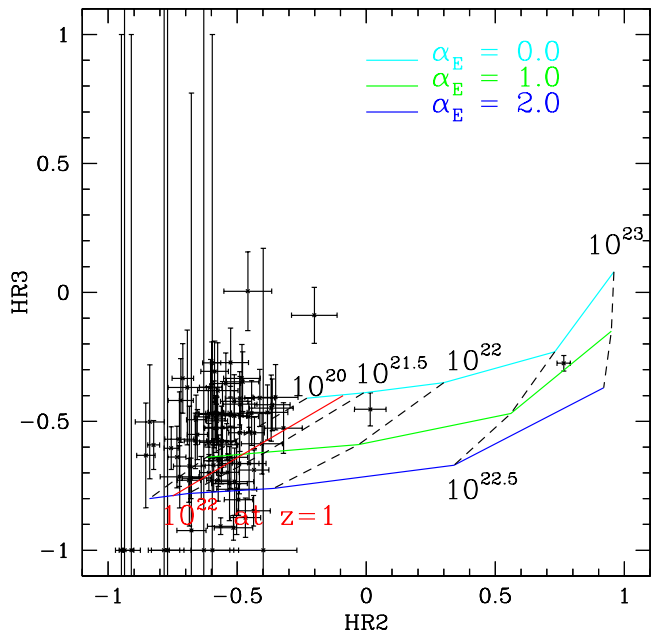, width=9cm}
  \end{center}
\caption{HR2 vs. HR3 for the unidentified objects. 
For clarity we have reported only the 85 unidentified 
sources with at least 100 counts in the 0.5-4.5 keV energy band.
We have also reported the expected HR2 and HR3 values for 
an absorbed power-law model as a function of $\alpha_E$ and 
$N_H$; the absorption has been assumed at $z\sim 0$.
The red line corresponds to an absorption with $N_H = 10^{22}$ cm$^{-2}$
at z=1. 
}  
\label{rdellaceca-WB1_fig:fig4}
\end{figure}

\begin{figure}[ht]
  \begin{center}
    \epsfig{file=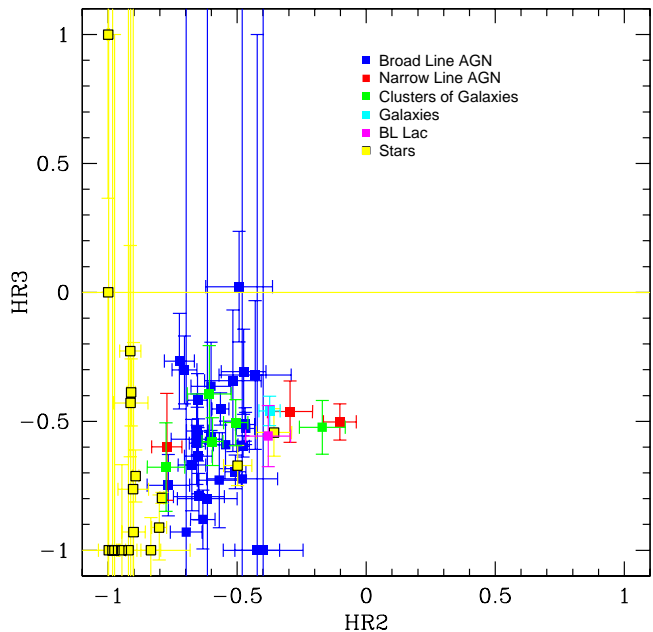, width=9cm}
  \end{center}
\caption{HR2 vs. HR3 for the identified objects.
We have used different colors to mark the identified objects. 
}  
\label{rdellaceca-WB1_fig:fig5}
\end{figure}

Combining the information on HR2 and HR3 we can study the broad 
band (0.5--7.5 keV) spectral properties of the sample as well as 
the selection function of the XMM-Newton BSS.  
In Figure~\ref{rdellaceca-WB1_fig:fig4} we report 
the position of the unidentified objects in the   
HR2-HR3 plane. Because of the large errors on HR3  for some of the
sources,  we have plotted here only the  85 unidentified sources with at least 
100 net counts in the 0.5--4.5 keV energy band.  We have also reported in this
figure the  expected HR2 and HR3 values for a simple  absorbed power-law model
as a function of $\alpha_E$ and  $N_H$.
In this simple model the absorption has been assumed at
$z\sim 0$; 
for comparison the red line corresponds to an absorption  
with $N_H= 10^{22}$ cm$^{-2}$at z=1.
Finally in Figure~\ref{rdellaceca-WB1_fig:fig5} 
we report the position of the identified objects in the 
HR2 - HR3 plane.

Useful information can be extracted by comparing 
the position of the sources with the ``locus" expected from a simple
absorbed power-law model.  The ``bulk" of the sources are strongly
clustered in the region where ``unabsorbed" objects are expected.
There are
very few objects with hardness ratios consistent with those
expected from an absorbed power-law model  having $N_H$ (at z=0) in the range
between $10^{21.5}$ and   $10^{23}$ cm$^{-2}$ (four more objects are
present in this area if we consider  also the unidentified
sources with less than 100 counts).
It is too early to speculate whether the paucity of these objects is due 
to selection effects (energy band, flux limit, etc..) or it is intrinsic. 
It will be instructive to compare the results
obtained from  the XMM-Newton BSS with the results obtained from other samples
selected at  harder energies e.g. in the 5--10 keV energy range.

Indeed we are able to detect absorbed objects;
for example we already have 3 Narrow Line AGN and two of these 
sources have the hardest spectra amongst the identified objects, 
highly suggestive of intrinsic absorption. 
It is worth noting that the class of ``Compton thick" absorbed AGN, 
which could represent a considerable fraction of the absorbed AGN population
(\cite{rdellaceca-WB1:Bassani1999}; \cite{rdellaceca-WB1:Risaliti1999}),    
could have X-ray spectral properties similar to that expected 
from unabsorbed AGN (see \cite{rdellaceca-WB1:Dellaceca1999}). 
Moreover 
an absorbed AGN at z=1 having  $N_H\sim 10^{22}$ cm$^{-2}$ should
have hardness ratios around the red line  reported in Figure~\ref{rdellaceca-WB1_fig:fig4}; 
at the flux
limit of the XMM-Newton BSS this AGN  should have an intrinsic X-ray luminosity  $\sim
3 \times 10^{44}$ erg s$^{-1}$ i.e. in the type 2 QSOs regime. 
Therefore, some Type
2 QSOs could be already present among the unidentified  sources. For these
objects the XMM-Newton BSS will be fundamental since, for many of  them, it will be
possible to compare their  X-ray spectral properties 
(e.g. presence of Fe lines, spectral slope, etc..) with the optical ones
(e.g. presence of broad and/or narrow line, colors etc..).

\section{Summary and Conclusions}

In this paper we have discussed ``The XMM-Newton Bright Serendipitous Source Sample"
(XMM-Newton BSS).  This sample is designed to contain 
$\sim 1000$ sources at high galactic latitude having  a flux limit of $\sim
10^{-13}$ erg cm$^{-2}$ s$^{-1}$  in the 0.5--4.5 keV energy band
\footnote {Given a density of about 1.8 XMM-Newton BSS source {\it per\/} EPIC MOS2 field 
we need about 
600 suitable EPIC MOS2 data set to build up the sample. At the actual XMM-Newton 
observation rate we expect to complete the selection and definition of 
the XMM-Newton BSS sample ($\sim 1000$ sources) in two years.}
. The main
scientific goals of the project are the study  of the mix of
the different kind of AGN,  the investigation of the cosmological evolution of
clusters of galaxies and the definition of a statistical identification
procedure to  select rare and interesting populations of X-ray sources using 
the ``huge" amount of data that will be accumulated by XMM-Newton in the next few years.

We have discussed the survey strategy with respect to the energy selection
band, primary selection camera (EPIC MOS2 detector),  criteria for field and source
selection. 

From the analysis of 104 EPIC MOS2 fields, a first  sample  of
185 X-ray sources
%,  having a count rate 0.5--4.5 keV $>= 1\times 10^{-2}$ cts/s, 
has been selected. 
%This count rate corresponds to a flux (0.5--4.5 keV)
%of  $\sim 7.2 \times 10^{-14}$ erg cm$^{-2}$ s$^{-1}$ (assuming  a power-law
%model with $\alpha_E = 0.9$, the mean energy spectral index  of the
%extragalactic population)  

A preliminary analysis indicates that:

a) the optical counterpart of the  majority (85-90\%) of these X-ray sources 
has an optical magnitude above the POSS II limit (R $\sim 21^{mag}$), thus 
allowing spectroscopic identification on a 4 meter telescope;

b) given the X-ray positional accuracy  of such bright sources (2-5 arcsec 
at the 90\% confidence level) and the magnitude of the optical counterpart only
one object needs  to be observed to obtain the optical identification; 

c) the majority of the X-ray sources have enough statistics (hundreds of 
counts if data from EPIC MOS1, MOS2 and pn are combined) to allow X-ray studies 
in terms of energy distributions, source extent and flux variability.

Fifty-nine sources (32\% of the sample) have been identified so far.
The large majority of the identified objects are broad line AGN (30); 
we have also found 3 narrow line AGN, 1 optically ``normal" galaxy, 
5 clusters of galaxies, 1 BL Lac object and 19 stars. However 
please note that the sample of identified objects is not statistically 
representative of the full sample.

Using the sample of 185 sources, we have investigated the number-flux 
relationship, which is well described by a power law model, N($>$S) = K $\times
S^{-\alpha}$, with best fit value for the slope of $\alpha = 1.65\pm 0.19$
($68\%$ confidence interval). A very good agreement with previous and new
results is obtained  making us confident of the correctness of data selection and 
analysis.

We have investigated the broad band spectral properties of the sample 
using the hardness  ratio method.  
No obvious trend of the source spectra as a function of the count rate
is clearly visible. The average spectra of the ``extragalactic"  population
corresponds to an energy spectral index of $\alpha_E = 0.9$; about 13\% of the
sources seem to be described  by an energy spectral index flatter
than that of the CXB ($\alpha_E = 0.4$).

A nice result is the finding of a clear separation, in the hardness ratio diagram, 
between galactic stars and extragalactic sources.

Using the position of the sources in the hardness ratio diagram
we have also discussed the possible composition of the XMM-Newton BSS as a function 
of intrinsic absorption. 
Apparently we found very few sources which seem to be described by 
an absorbed power-law having $N_H$ (at z=0) in the range
between $10^{21.5}$ and   $10^{23}$ cm$^{-2}$.
%a result in part expected given the energy selection band. 
To shed light on the CXB modeling
it will be important to compare the source composition of this sample with 
the source composition of complete samples selected at harder energies 
(e.g. 5--10 keV energy range). 
On the other hand our sample could contain ``Compton thick" local 
sources and/or high luminosity 
type 2 AGN. In this respect the XMM-Newton BSS will be, once fully 
identified, an important tool  
to investigate and to compare in detail X-ray and 
optical/infrared spectral properties.  

Finally the XMM-Newton BSS will be instrumental  to investigating the cosmological 
evolution properties of clusters of galaxies. Up to now only 5  X-ray sources
have been identified as clusters of galaxies.  On the basis of known
cosmological properties we expect between  120 (in the case of negative
evolution) and 160 (in the case of no evolution)  clusters of galaxies; the
largest discrepancy is on the predicted  number of clusters in the high
luminosity - high z plane, a region not well sampled yet. 
An appropriate source detection algorithm for extended sources 
(e.g. wavelet) is under investigation.

\begin{acknowledgements}
This work is based partly on observations with XMM-Newton, 
an ESA Science Mission with instruments and contributions
directly funded by ESA Members States and the USA (NASA). 
The XMM-Newton project is supported by the Bundesministerium f\"ur Bildung 
und Forschung / Deutsches Zentrum f\"ur Luft- und Raumfahrt (BMBF/DLR), the 
Max-Planck Society and the Heidenhain-Stiftung.
This research has made use of the NASA/IPAC Extragalactic
Database (NED; which is operated by the Jet Propulsion Laboratory, 
California Institute of Technology, under contract with the National 
Aeronautics and Space Administration) and of the SIMBAD database 
(operated at CDS, Strasbourg, France).
RDC, TM and AC acknowledge partial financial support by the 
Italian Space Agency (ASI) and by the
MURST (Cofin00-32-36). PS acknowledge partial financial support 
by the Italian {\it Consorzio Nazionale per l'Astronomia e l'Astrofisica} 
(CNAA).
XB and FJC acknowledge financial support from the Spanish MCyT under 
project AYA2000-1690.
AS was supported by the DLR under grant 50 OX 9801 3.
\end{acknowledgements}


\begin{thebibliography}{}

\bibitem[\protect\astroncite{Baldi et~al.}{2002}]{rdellaceca-WB1:Baldi2002}
Baldi, A., Molendi, S., Comastri, A., Fiore, F., Matt, G., \& Vignali, C.\ 2002, Ap.J., 564, 190

\bibitem[\protect\astroncite{Barcons et~al.}{2002}]{rdellaceca-WB1:Barcons2002}
Barcons, X., Carrera, F.G., Watson, M.G., et~al., 2002, A\&A, 382, 522

\bibitem[\protect\astroncite{Bassani et~al.}{1999}]{rdellaceca-WB1:Bassani1999}
Bassani, L., Dadina, M., Maiolino, R., Salvati, M., Risaliti, G., Della Ceca, R., 
Matt, G., \& Zamorani, G.\ 1999, Ap.J. Supp., 121, 473

\bibitem[\protect\astroncite{Borgani et~al.}{2001}]{rdellaceca-WB1:Borgani2001}
Borgani, S.~et al.\ 2001, Ap.J, 561, 13

\bibitem[\protect\astroncite{Condon et~al.}{1998}]{rdellaceca-WB1:Condon1998}
Condon, J.~J., Cotton, W.~D., Greisen, E.~W., Yin, Q.~F., Perley, R.~A., Taylor, G.~B., 
\& Broderick, J.~J.\ 1998, A.J., 115, 1693

\bibitem[\protect\astroncite{Della Ceca et~al.}{1999}]{rdellaceca-WB1:Dellaceca1999}
Della Ceca, R., Castelli, G., Braito, V., Cagnoni, I., \& Maccacaro, T.\ 1999, Ap.J., 524, 674

\bibitem[\protect\astroncite{Della Ceca et~al.}{2000}]{rdellaceca-WB1:Dellaceca2000}
Della Ceca, R., Maccacaro, T., Rosati, P., \& Braito, V.\ 2000, A\&A, 355, 121

\bibitem[\protect\astroncite{Ehle et~al.}{2001}]{rdellaceca-WB1:Ehle2001}
Ehle, M., et~al.\ 2001, XMM-Newton Users' Handbook

\bibitem[\protect\astroncite{Gioia et~al.}{1990}]{rdellaceca-WB1:Gioia1990}
Gioia, I.~M., Maccacaro, T., Schild, R.~E., Wolter, A., Stocke, J.~T., Morris, S.~L., \& 
Henry, J.~P.\ 1990, Ap.J. Supp., 72, 567

\bibitem[\protect\astroncite{Hasinger et~al.}{1998}]{rdellaceca-WB1:Hasinger1998}
Hasinger, G., Burg, R., Giacconi, R., Schmidt, M., Trumper, J., 
\& Zamorani, G.\ 1998, A\&A, 329, 482

\bibitem[\protect\astroncite{Hasinger et~al.}{2001}]{rdellaceca-WB1:Hasinger2001}
Hasinger, G., et~al.\ 2001, A\&A, 365, L45

\bibitem[\protect\astroncite{Maiolino et~al.}{2001}]{rdellaceca-WB1:Maiolino2001}
Maiolino, R., Marconi, A., Salvati, M., Risaliti, G., Severgnini, P., Oliva, E., 
La Franca, F., \& Vanzi, L.\ 2001, A\&A, 365, 28

\bibitem[\protect\astroncite{Miyaji et~al.}{(2000)}]{rdellaceca-WB1:Miyaji2000}
Miyaji, T., Hasinger, G.~;., \& Schmidt, M.\ 2000, A\&A, 353, 25

\bibitem[\protect\astroncite{Risaliti et~al.}{1999}]{rdellaceca-WB1:Risaliti1999}
Risaliti, G., Maiolino, R., \& Salvati, M.\ 1999, Ap.J, 522, 157

\bibitem[\protect\astroncite{Rosati et~al.}{1998}]{rdellaceca-WB1:Rosati1998}
Rosati, P., Della Ceca, R., Norman, C., \& Giacconi, R.\ 1998, Ap.J. Lett., 492, L21

\bibitem[\protect\astroncite{Watson et~al.}{2001}]{rdellaceca-WB1:Watson2001}
Watson, M.~G.~et al.\ 2001, A\&A, 365, L51


\end{thebibliography}
\end{document}